\documentclass[aps,prd,amsmath,amssymb,preprintnumbers,superscriptaddress,nofootinbib,a4paper,11pt]{revtex4-1}
\pdfoutput=1
\usepackage{amsthm}
\usepackage{graphicx}
\usepackage[ margin=5pt, font=small,labelfont=bf, justification=raggedright]{caption}
\usepackage{url}
\usepackage[bookmarks, pagebackref=false]{hyperref}
\usepackage{color}
	\definecolor{rossoCP3}{cmyk}{0,.88,.77,.40}
		\definecolor{graa}{rgb}{0.8,0.8,0.8}
		\definecolor{blaa}{rgb}{0.2,0.2,0.6}
		\hypersetup{
			colorlinks,
			bookmarksopen,
			bookmarksnumbered,
			citecolor=blaa, 		
			linkcolor=rossoCP3,	
			urlcolor=rossoCP3,			
			}
\usepackage{dcolumn}
\usepackage{bm}
\usepackage{bbm}
\usepackage{pxfonts}

\usepackage{amsmath}
\usepackage{amsfonts}

\usepackage{dcolumn}
\usepackage{bm}
\usepackage{amssymb}
\usepackage{latexsym}
\usepackage{color}

\usepackage{epsfig}
\usepackage{placeins}

\usepackage{youngtab}
\usepackage{slashed}
\usepackage{subfig}

\begin{document}

\title{\Large  Effect of four-fermion operators \\ on the mass of the composite particles}
\author{Roshan Foadi}
\email{roshan.foadi@gmail.com}

\begin{abstract}
We propose a theoretical framework for evaluating the effect of four-fermion operators on the mass of composite particles in confining strongly-coupled gauge theories. The confining sector is modelled by a non-local Nambu-Jona Lasinio action, whereas the four-fermion operators, arising from a different sector, are local. In order to illustrate the method, we investigate a simple toy model with a global $SU(2)_L\times SU(2)_R\to SU(2)_V$ symmetry breaking, and a four-fermion operator breaking $SU(2)_L\times SU(2)_R$ but preserving $SU(2)_V$. In the particle spectrum we only include the pseudoscalar isospin triplet, that is the pseudo-Nambu-Goldstone bosons associated with chiral symmetry breaking, and the lightest scalar singlet. After checking that the nonlocal model successfully accounts for the experimental results in two-flavour QCD, we investigate the mass spectrum as a function of the four-fermion coupling. For our specific choice of four-fermion operator, we find that the mass of the  pseudoscalar triplet grows, whereas the mass of the lightest scalar singlet is approximately unaffected, as the four-fermion coupling grows. We argue that these results can be directly tested on the lattice, and briefly discuss possible applications of this technique to models of dynamical electroweak symmetry breaking.  
\end{abstract}

\maketitle

\section{Introduction}
In recent years strongly coupled gauge theories (SCGTs) have drawn great interest because of their possible connection with the mechanism of dynamical electroweak symmetry breaking (DEWSB). Any realistic model of DEWSB must feature, at the electroweak scale, a set of four-fermion operators (4FOs) , which arise from a higher energy scale by integrating out massive gauge bosons. Such 4FOs are in fact the source of the Standard-Model fermion masses \cite{Dimopoulos:1979es,Eichten:1979ah,Kaplan:1991dc,Hill:2002ap}, and play a major role in determining the mass of the lightest scalar resonance, which is to be identified with the Higgs boson. The latter, for instance, the can be the pseudo-Nambu-Goldstone boson (pNBG) of a global symmetry, broken by the vacuum state of the SCGT \cite{Kaplan:1983sm,Agashe:2004rs,Cacciapaglia:2014uja}. In this case 4FOs are necessary to give the Higgs a potential, and since the largest 4FOs are those connected with the generation of the top mass, this leaves us with the intriguing possibility that the Higgs and top mass are related \cite{Agashe:2004rs,Cacciapaglia:2014uja}. Alternatively, it has been argued that the lightest scalar resonance of a SCGT receive negative radiative corrections from 4FOs, which may lower its mass from $\lesssim 1$ TeV to 125 GeV \cite{Foadi:2012bb,DiChiara:2014gsa,DiChiara:2014uwa}. Furthermore 4FOs may increase the mass of other pNGBs which the SCGT may feature, explaining why these have thus far  evaded detection \cite{Hill:2002ap}. It is therefore evident that investigating the effect of 4FOs on the mass spectrum of SCGTs is of utmost importance in the context of DEWSB.

In this note we shall not investigate any model of DEWSB, but rather use a simple toy model to illustrate the method for computing the effect of 4FOs on the mass spectrum of a confining SCGT. Specifically, we take the confined fermions $\Psi_{i\, L}\equiv (U_{i\, L},D_{i\, L})$ and $\Psi_{i\, R}\equiv (U_{i\, R},D_{i\, R})$ in a complex representation $R$ of the gauge group, where $i=1\dots N\equiv{\rm dim}(R)$ is the gauge index. This theory features a global $SU(2)_L\times SU(2)_R$ symmetry, which is broken by the vacuum state to $SU(2)_V$. As a consequence, an $SU(2)_V$ triplet of pseudoscalar massless NGBs is generated. This is an upper scale version of the QCD pion triplet, which is massless in the chiral limit. Pursuing the QCD analogy further, we expect the lightest massive resonance to be a scalar singlet, {\em i.e.} a scaled-up sigma meson with a mass of order of the confinement scale  $\kappa$. We then include the 4FO $(\overline{\Psi}\Psi)^2$: this is assumed to be generated by integrating out a gauge boson with mass ${\cal M}$, where ${\cal M}^2\gg \kappa^2$. Such an operator preserves $SU(2)_V$ but breaks $SU(2)_L\times SU(2)_R$, thus generating a mass for the pNGBs. Our goal is to compute the latter and the mass of the scalar singlet. In order to achieve this, we employ the Nambu--Jona-Lasinio (NJL) approximation of the SCGT, and compute correlation functions in the large-$N$ limit.

\begin{figure}
\includegraphics[width=3.5in]{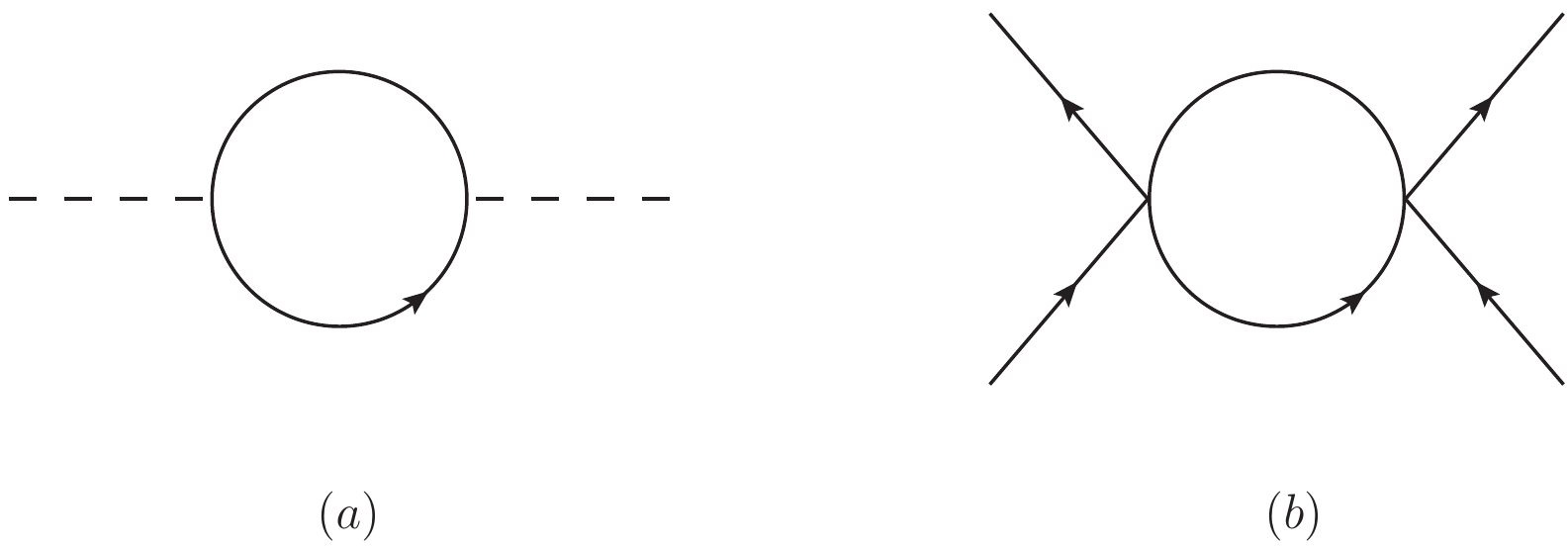}
\caption{(a) Hadronization diagram: the loop consists of constituent fermions, whereas the external particles are composite states. (b) Renormalization diagram: both loop and external particles are constituent fermions, and this diagram amount to a renormalization of the four-fermion operator.}
\label{Fig:YesNoHadron}
\end{figure}

The standard NJL models with a cutoff fail to account for confinement, and are therefore not fully reliable. In this note we avoid this issue by employing a nonlocal Nambu--Jona-Lasinio model (nlNJL) \cite{Diakonov:1985eg,Bowler:1994ir,Plant:1997jr,Anikin:2000rq,Volkov:2006vq}. This accounts for confinement through a constituent fermion mass which is exponentially suppressed at large momenta, leading to a propagator without pole singularities. Furthermore, the same exponential damping makes hadronization diagrams, such as the one of Fig. \ref{Fig:YesNoHadron} (a), finite. Diagrams without hadronization into external or virtual composite states are those only involving the four-fermion vertex, such as the diagram of  Fig. \ref{Fig:YesNoHadron} (b). These are not finite, and must be cutoff at the 4FO mass scale ${\cal M}$.

The most interesting aspect of this model is that it is predictive, and can therefore be tested with computations in lattice gauge theory. Of course on the lattice one would not employ a nlNJL approximation of the SCGT, but rather use the full gauge theory, and augment the latter with a local 4FO. An example of this type of analysis is provided by the lattice study of gauged NJL models \cite{Catterall:2011ab,Catterall:2013koa}, with 4FOs which are themselves local NJL Lagrangians.

In the remainder of this note we illustrate the model by computing pseudoscalar mass and decay constant, as well as the scalar mass. We then discuss the results, and briefly comment on the application of this technique to theories of DEWSB.

\section{The non-local NJL model}
Using a linear representation of the chiral symmetry, and suppressing gauge as well as isospin indices, the NJL Lagrangian for our model reads
\begin{equation}
{\cal L} = \overline{\Psi}\, i\, \slashed{\partial} \Psi-\frac{\Lambda^2}{2}\left[S^{\prime 2}+(\Pi^a)^2\right]-J_S\, S^\prime - J_\Pi^a\, \Pi^a + \frac{g^2}{2{\cal M}^2}\left(\overline{\Psi}\Psi\right)^2\ ,
\label{Eq:L}
\end{equation}
where $\Pi^a$ is the pNGB pseudoscalar triplet and $S^\prime$ is a scalar singlet. The latter acquires a nonzero vacuum expectation value from fermion condensation. This prompts us to define
\begin{equation}
S^\prime = \kappa + S\ ,
\label{Eq:Shift}
\end{equation}
where $S$ has zero vacuum expectation value, and $\kappa$ is the confinement scale. Note that we have introduced three different mass scales: $\Lambda$, $\kappa$ and ${\cal M}$. We shall see that $\Lambda$ is fixed by the requirement that there is no linear term in the scalar field $S$, and is related to the confinement scale $\kappa$. The 4FO scale ${\cal M}$ is a new independent scale, completely unrelated to $\Lambda$ and $\kappa$. However for our Lagrangian to be meaningful, we must have ${\cal M}^2\gg \kappa^2,\Lambda^2$, or else a 4FO is no longer a good approximation for the physics at the scale ${\cal M}$.

The nlNJL model we employ is based on the instanton vacuum, and has been successfully applied to low-energy QCD. The nonlocal currents are
\begin{equation}
J_I(x) = \int d^4 x_1 \int d^4 x_2\, f(x_1) f(x_2)\, \overline{\Psi}(x-x_1)\Gamma_I \Psi(x-x_2)\ ,
\end{equation}
where $\Gamma_S=1$, $\Gamma_\Pi=i\, \gamma_5\, \tau^a$, and $\tau^a$ are the Pauli matrices. The function $f(x)$ must be chosen to insure fermion confinement and to make all integrals with external $\Pi$ or $S$ finite. In the chiral limit, $g\to 0$, this can be achieved by choosing the nonsingular fermion propagator
\begin{equation}
\frac{1}{P^2+M_{P^2}^2}  \underset{g\to 0}{\longrightarrow} \frac{1-\exp(-P^2/\kappa^2)}{P^2} \ .
\end{equation}
Here and below we mostly use Euclidean momenta, $P^0 \equiv i\, p^0\ ,  P^i \equiv p^i $, which we always denote with capital letters. Solving for the momentum-dependent fermion mass gives
\begin{equation}
M_{P^2}^2  \underset{g\to 0}{\longrightarrow} \frac{P^2}{\exp(P^2/\kappa^2)-1} \ .
\end{equation}
We see that the confinement scale $\kappa$ is nothing but the fermion mass at zero momentum. Let $f_{P^2}$ be the Fourier transform of $f(x)$. In the chiral limit, this is related to the momentum-dependent fermion mass by 
\begin{equation}
M_{P^2} \underset{g\to 0}{\longrightarrow} \kappa\, f_{P^2}^2\ ,
\end{equation}
whence
\begin{equation}
f_{P^2} = \left[\frac{P^2/\kappa^2}{\exp(P^2/\kappa^2)-1}\right]^{1/4} \ .
\end{equation}
Note that this model breaks down at large time-like momenta, as Green functions grow exponentially, and unphysical poles appear \cite{Plant:1997jr}. Therefore, this model is inadequate to account for radial excitations, which require a different treatment \cite{Volkov:1999yi}. 

We now let $g\neq 0$, and compute the momentum-dependent fermion mass, the pseudoscalar mass and decay constant, and the scalar mass. As we compute these quantities in the large-$N$ scheme, we take the four-fermion coupling $g$ to scale like $1/\sqrt{N}$.
\begin{figure}
\includegraphics[width=4.5in]{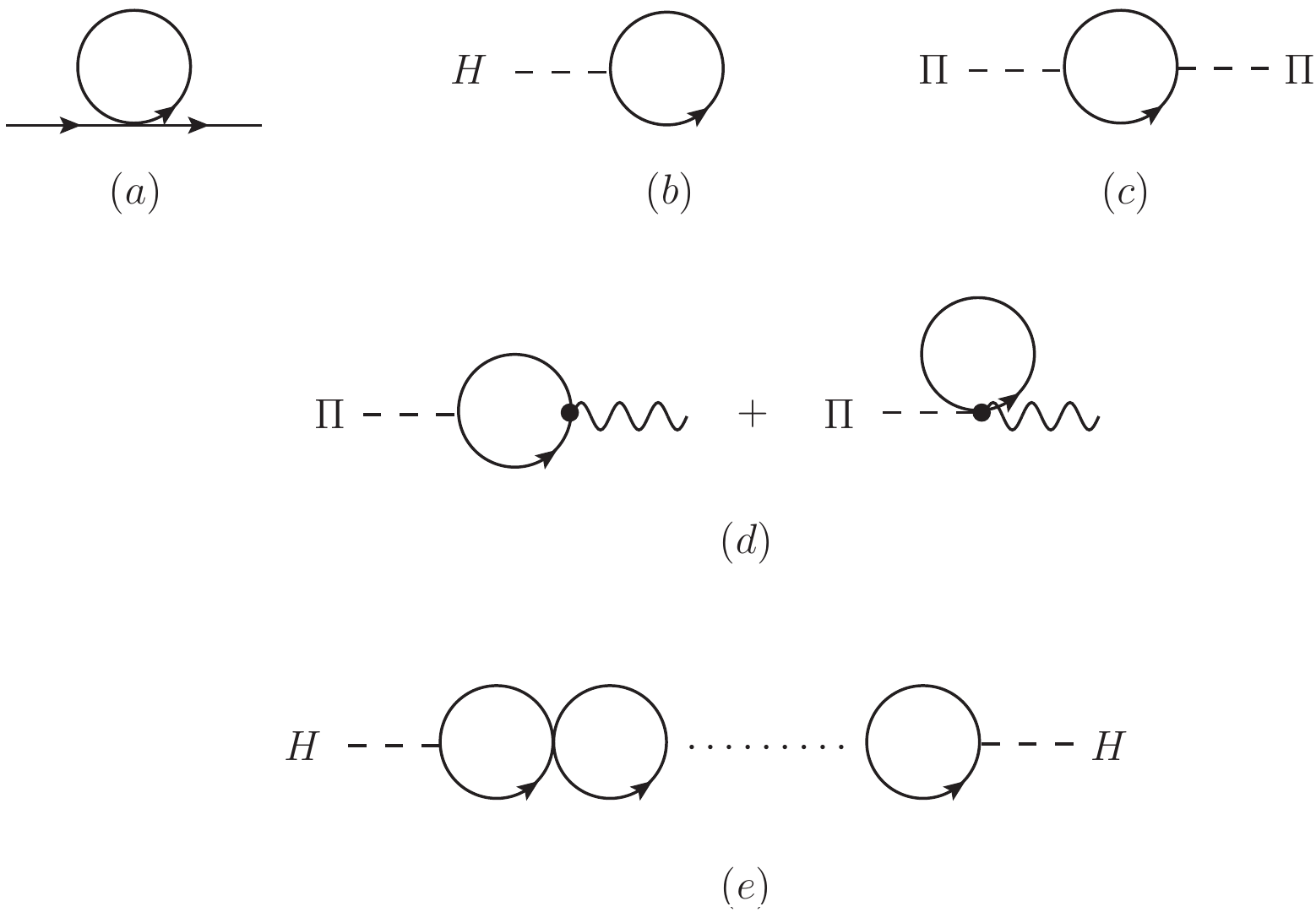}
\caption{Diagrams contributing to: (a) Fermion mass, (b) H linear term (to be cancelled by the tree-level contribution), (c) pseudoscalar self-energy, (d) pseudoscalar decay constant, and (e) scalar self-energy.}
\label{Fig:Diagrams}
\end{figure}
\subsection{Fermion mass}
To leading order the fermion mass receives a momentum-independent contribution through the diagram of Fig. \ref{Fig:Diagrams} (a). This gives the mass
\begin{equation}
M_{P^2} = \kappa\, f_{P^2} + m\ ,
\end{equation}
where $m$ is found by solving the integral equation
\begin{equation}
m =\frac{8\, N\, g^2}{{\cal M}^2}\int \frac{d^4 K}{(2\pi)^4}\frac{\kappa\, f_{K^2} + m}{K^2+M_{K^2}^2}\ .
\label{Eq:m}
\end{equation}
A few aspects of this computation need to be clarified. First, $m$ momentum independence is only an approximation valid for momenta below ${\cal M}$. Second, the second term in the numerator of the integrand multiplies no damping factor, and the corresponding integral must be cutoff at the scale ${\cal M}$. This brings in a dependence on ${\cal M}$ which is different from the one introduced by the 4F coupling. Finally, a nonzero $m$ brings in singularities in the complex $K^2$ plane. This does not spoil confinement (as the singularities are not on the real axis and no free-fermion asymptotic states are possible) but may prevent us from Wick-rotating the integration contour to the real $K^0$ axis. However, for small values of $m/\kappa$ we can expand the integrand, and each term of the expansion is free of singularities. Then we can Wick-rotate, sum the series, and analytically continue to larger values of $m/\kappa$. This woks for the fermion mass above, but also for the computation of the pseudoscalar self-energy and decay constant, as well as the scalar self-energy. 
\subsection{Linear term cancellation}
The shift introduced in (\ref{Eq:Shift}) is to cancel the tadpole diagram of Fig. \ref{Fig:Diagrams} (b). This imposes the relationship
\begin{equation}
\Lambda^2 = 8\, N\, I
\label{Eq:LTC}
\end{equation}
where
\begin{equation}
I \equiv \frac{1}{\kappa}\int  \frac{d^4 K}{(2\pi)^4} \frac{f_{K^2}^2\, M_{K^2}}{K^2+M_{K^2}^2}\ .
\end{equation}
In the chiral limit, $g\to 0$, (\ref{Eq:LTC}) becomes
\begin{equation}
\Lambda^2  \underset{g\to 0}{\longrightarrow} \frac{N}{2\pi^2}\, \kappa^2 \ ,
\end{equation}
explicitly showing the relation between $\Lambda$ and $\kappa$.
\subsection{Pseudoscalar mass and decay constant}
To leading order in $1/N$ the pseudoscalar self-energy is given by the tree-level mass term plus the diagram of Fig. \ref{Fig:Diagrams} (c), where the former is fixed by condition (\ref{Eq:LTC}). This leads to the amplitude
\begin{equation}
\Pi_\Pi(Q^2) = 8\, N\left[-I+I^{PP}(Q^2)\right]\ ,
\label{Eq:PP}
\end{equation}
where
\begin{equation}
I^{PP}(Q^2)=\int  \frac{d^4 K}{(2\pi)^4}\frac{f_{K^2}^2\, f_{(K+Q)^2}^2\Big(K\cdot(K+Q)+ M_{K^2}\, M_{(K+Q)^2}\Big)}{\big(K^2+M_{K^2}^2\big)\big((K+Q)^2+M_{(K+Q)^2}^2\big)} \ .
\end{equation}
The pseudoscalar mass can be found by solving numerically the equation
\begin{equation}
\Pi_\Pi(-M_\Pi^2) = 0\ .
\label{Eq:MassMP}
\end{equation}
For small values of $m$ and $Q^2$, (\ref{Eq:PP}) becomes
\begin{equation}
\Pi_\Pi(Q^2) = -  \frac{1}{\kappa^2}\, m\, \langle\overline{\Psi}\Psi\rangle - Z_\Pi^{-1} Q^2 + {\cal O}(Q^4) + {\cal O}(m^2)\ ,
\end{equation}
where $Z_\Pi^{-1}$ is the pseudoscalar wavefunction renormalization,
\begin{equation}
Z_\Pi^{-1} \equiv - \Pi_\Pi^\prime(-M_\Pi^2)\ ,
\end{equation}
and $\langle\overline{\Psi}\Psi\rangle$ is the fermion condensate,
\begin{equation}
\langle\overline{\Psi}\Psi\rangle = 8\, N \int  \frac{d^4 K}{(2\pi)^4} \frac{M_{K^2}}{K^2+M_{K^2}^2}\ .
\end{equation}
This shows that in the chiral limit the self-energy vanishes at zero external momenta, proving that the pseudoscalar triplet is an exact NGB in absence of chiral symmetry violations. 

In order to compute the decay constant we need an expression for the axial current. This is found as usual by requiring the action, in the chiral limit $g\to 0$, to be invariant under a local axial transformation,
\begin{align}
& \delta_A \Psi(x) = -i\, \delta\epsilon_A^a(x)\, \gamma_5\, T^a\, Q(x)\ , \nonumber \\
& \delta_A \overline{\Psi}(x)  = -i\, \delta\epsilon_A^a(x)\, \overline{\Psi}(x)\, \gamma_5\, T^a\ , \nonumber \\
& \delta_A S^\prime(x)  = -\delta\epsilon_A^a(x)\, \Pi^a(x)\ , \nonumber \\
& \delta_A \Pi^a(x)  = \delta\epsilon_A^a(x)\, S^\prime(x)\ ,
\end{align}
where $2T^a=\tau^a$. This leads to the standard local piece plus a nonlocal piece:
\begin{align}
&-\frac{\delta}{\delta\epsilon_A^a(x)}\int d^4 y\, {\cal L}
= \frac{\partial}{\partial x^\mu} \overline{\Psi}(x)\, \gamma^\mu\, \gamma_5\, T^a\, \Psi(x)  \nonumber \\
& - \int d^4 y \int d^4 y_1 \int d^4 y_2 \, f(y_1)\, f(y_2) \Bigg\{\Big(2\delta^4(y-x)-\delta^4(y-y_1-x)-\delta^4(y-y_2-x)\Big) \nonumber \\
&\times \left[ i\, (\kappa+S(y))\, \overline{\Psi}(y-y_1)\, \gamma_5 T^a\, \Psi(y-y_2) - \frac{\Pi^a(y)}{2} \overline{\Psi}(y-y_1)\, \Psi(y-y_2) \right] \nonumber \\
& +i\, \epsilon^{a b c} \Pi^b(y) \Big(\delta^4(y-y_1-x)-\delta^4(y-y_2-x)\Big) \overline{\Psi}(y-y_1)\, T^b\, \Psi(y-y_2) \Bigg\} = 0 \ .
\end{align}
We can express the nonlocal piece as a total divergence by using the identity
\begin{equation}
\delta^4(x-x_A) - \delta^4(x-x_B) = \int_0^1 d\lambda \frac{d z^\mu}{d\lambda}\, \frac{\partial}{\partial^\mu}\delta^4(z-x)\ ,
\end{equation}
where $z(\lambda)$ is an arbitrary path with initial point $z(0)=x_A$ and endpoint $z(1)=x_B$ \cite{Bowler:1994ir}. For a straight line,
\begin{equation}
z(\lambda) = (1-\lambda)\, x_A + \lambda\, x_B\ ,
\end{equation}
this leads to the conserved current
\begin{align}
& j_A^{a\mu}(x) = \overline{\Psi}(x)\, \gamma^\mu\, \gamma_5\, T^a\, \Psi(x) \nonumber \\
& -\int_0^1 d\lambda \int d^4 y \int d^4 y_1 \int d^4 y_2\, f(y_1)\, f(y_2)\Bigg\{
\left[y_1^\mu\, \delta^4(x-y+\lambda\, y_1)+y_2^\mu\, \delta^4(x-y+\lambda\, y_2)\right] \nonumber \\
& \times\left[i\, (v+S(y))\, \overline{\Psi}(y-y_1)\, \gamma_5\, T^a\, \Psi(y-y_2) - \frac{\Pi^a(y)}{2} \overline{\Psi}(y-y_1)\, \Psi(y-y_2)\right] \nonumber \\
&  + (y_2-y_1)^\mu\, \delta^4\left[(1-\lambda)\, y_1+\lambda\, y_2-y+x\right]\, i\, \epsilon^{abc}\, \Pi^b\, \overline{\Psi}(y-y_1)\, T^c\, \Psi(y-y_2)\Bigg\} \ .
\label{Eq:jA}
\end{align}
This current is not uniquely defined, as we can choose a different path to express the difference between delta functions. However the longitudinal component is unique, and the latter is the one needed to evaluate the pseudoscalar decay constant $F_\Pi$: 
\begin{equation}
\langle 0 | q_\mu j_A^{a\mu}(q) |\Pi^a(q) \rangle  \underset{q^2\to M_\Pi^2}{\longrightarrow} -i\, \delta^{ab} F_\Pi\, q^2
\label{Eq:DefFP}
\end{equation}
Taking the Fourier transform of (\ref{Eq:jA}) and extracting the vertices, allows us to compute the matrix element on the left-hand side. To leading order in $1/N$ the latter is given by the diagrams of Fig. \ref{Fig:Diagrams} (d). Using Euclidean momenta, this eventually leads to the expression
\begin{equation}
F_\Pi = 4\, N\, Z_\Pi^{1/2}\, \frac{{\cal I}(-M_\Pi^2)-{\cal I}^{PA}(-M_\Pi^2)}{M_\Pi^2}\ ,
\label{Eq:FP}
\end{equation}
where
\begin{align}
& {\cal I}(Q^2) \equiv \int  \frac{d^4 K}{(2\pi)^4} \frac{\Big(f_{(K+Q)^2}\, f_{K^2}+f_{(K-Q)^2}\, f_{K^2}-2f_{K^2}^2\Big)\, M_{K^2}}{K^2+M_{K^2}^2} \ , \\
& {\cal I}^{PA}(Q^2) \equiv \int  \frac{d^4 K}{(2\pi)^4} \frac{M_{K^2}\, Q^2+K\cdot Q\, \Big(f_{K^2}^2-f_{(K+Q)^2}^2\Big)
+\Big(f_{K^2}-f_{(K+Q)^2}\Big)^2\Big(K\cdot(K+Q)+ M_{K^2}\, M_{(K+Q)^2}\Big)}{\big(K^2+M_{K^2}^2\big)\big((K+Q)^2+M_{(K+Q)^2}^2\big)} \ .
\end{align}
Note that $Z_\Pi^{1/2}$ scales like $1/\sqrt{N}$, and thus $F_\Pi$ properly scales like $\sqrt{N}$. For small values of $m$ and $q^2\equiv-Q^2$ the integrals above lead to
\begin{equation}
\langle 0 | q_\mu j_A^{a\mu}(q) |\Pi^a(q) \rangle = -i\, \delta^{ab}\, q^2\,  Z_\Pi^{-1/2}\, \kappa  + {\cal O}(q^4) + {\cal O}(m) \ ,
\end{equation}
whence
\begin{equation}
F_\Pi = Z_\Pi^{-1/2} \kappa + {\cal O}(m)\ .
\end{equation}
In this limit we can also solve (\ref{Eq:PP}) to obtain the pseudoscalar mass. This, together with the expression for $F_\Pi$, readily leads to the Gell-Mann--Oakes--Renner formula \cite{GellMann:1968rz,Bowler:1994ir}:
\begin{equation}
F_\Pi^2\, M_\Pi^2 = m\, \langle\overline{\Psi}\Psi\rangle + {\cal O}(m^2)\ .
\end{equation}
\subsection{Scalar mass}
The self-energy of the scalar singlet receives direct contribution from the four-fermion operator, and to leading order in $1/N$ we need to sum the chain of diagrams shown in Fig. \ref{Fig:Diagrams} (e). Including the tree-level mass term, which is fixed by condition (\ref{Eq:LTC}), this leads to the expression
\begin{equation}
\Pi_S(Q^2) = 8\, N\left[-I+I^{SS}(Q^2)+\frac{\displaystyle{\frac{8\, N\, g^2}{{\cal M}^2}}\Big(J^{SS}(Q^2)\Big)^2}{1-\displaystyle{\frac{8\, N\, g^2}{{\cal M}^2}}\, K^{SS}(Q^2)}\right] \ ,
\label{Eq:PSchain}
\end{equation}
where
\begin{align}
& I^{SS}(Q^2)=\int  \frac{d^4 K}{(2\pi)^4}\frac{f_{K^2}^2\, f_{(K+Q)^2}^2\Big(K\cdot(K+Q) - M_{K^2}\, M_{(K+Q)^2}\Big)}{\big(K^2+M_{K^2}^2\big)\big((K+Q)^2+M_{(K+Q)^2}^2\big)} \ , \\
& J^{SS}(Q^2)=\int  \frac{d^4 K}{(2\pi)^4}\frac{f_{K^2}\, f_{(K+Q)^2}\Big(K\cdot(K+Q) - M_{K^2}\, M_{(K+Q)^2}\Big)}{\big(K^2+M_{K^2}^2\big)\big((K+Q)^2+M_{(K+Q)^2}^2\big)} \ , \\
& K^{SS}(Q^2)=\int  \frac{d^4 K}{(2\pi)^4}\frac{K\cdot(K+Q) - M_{K^2}\, M_{(K+Q)^2}}{\big(K^2+M_{K^2}^2\big)\big((K+Q)^2+M_{(K+Q)^2}^2\big)} \ .
\end{align}
The last integral corresponds to a diagram like the one of Fig. \ref{Fig:YesNoHadron} (b), and must be cutoff at the scale ${\cal M}$. The scalar mass is obtained by solving numerically the equation
\begin{equation}
\Pi_S(-M_S^2) = 0\ .
\label{Eq:PS}
\end{equation}
\subsection{Two-flavour QCD}
With respect to the pseudoscalar mass and decay constant, the only effect of the four-fermion operator $(\overline{\Psi}\Psi)^2$ is to introduce an effective current fermion mass $m$, which adds to the momentum-dependent constituent mass. In two-flavour QCD the former should be close to $m_{u,d}\simeq 10\ {\rm MeV}$, the up- and down-quark current mass at the QCD scale. Requiring the pion mass $m_\pi\simeq 135$ MeV and the pion decay constant $f_\pi\simeq 92.8$ MeV, gives a unique prediction for  
$\kappa_{\rm QCD}$ and $m_{u,d}$:
\begin{equation}
\kappa_{\rm QCD} \simeq 330\ {\rm MeV}\ , \quad
m_{u,d} \simeq 12\ {\rm MeV} \ .
\end{equation}
The value of $m_{u,d}$ is sightly larger than it should be, but we should keep in mind that this is a leading order prediction in $1/N$, and $m_{u,d}$ is a rather sensitive quantity. A 7\% subleading correction to the pion mass, for instance, lowers $m_{u,d}$ to $\sim$10 MeV.

The lightest scalar singlet in QCD is the $\sigma$ meson, with a measured mass around $450\ {\rm MeV}$ \cite{GarciaMartin:2011jx}. Our model gives a direct correction from the 4FO to the scalar mass, and thus we cannot use the full result as a prediction for $m_\sigma$. However, removing the direct corrections from the 4FO, {\em i.e.} diagrams with chains of two or more loops, yields the correct prediction for $m_\sigma$. Computation gives
\begin{equation}
m_\sigma \simeq 425\ {\rm MeV}\ .
\label{Eq:msigma}
\end{equation}
This is in remarkable agreement with the experimental results, especially considering that it is a leading-order prediction, and is thus subject to $\sim$ 10\% corrections. For comparison, the local-NJL model prediction for $m_\sigma$ is 550 MeV \cite{Volkov:2006vq}. 

These results suggest that nlNJL models can successfully account for low-energy SCGTs which, similarly to QCD, are precociously asymptotically free. On the other hand, it is reasonable to expect nlNJL models to be inadequate for describing SCGTs with walking dynamics. In fact, while the former are characterised by a single mass scale, {\em i.e.} the confinement scale, the latter feature two hierarchically distinct mass scales, {\em i.e.} the confinement scale and the scale at which the near conformal behaviour is lost.
\subsection{Results}
\begin{figure}
\includegraphics[width=2.9in]{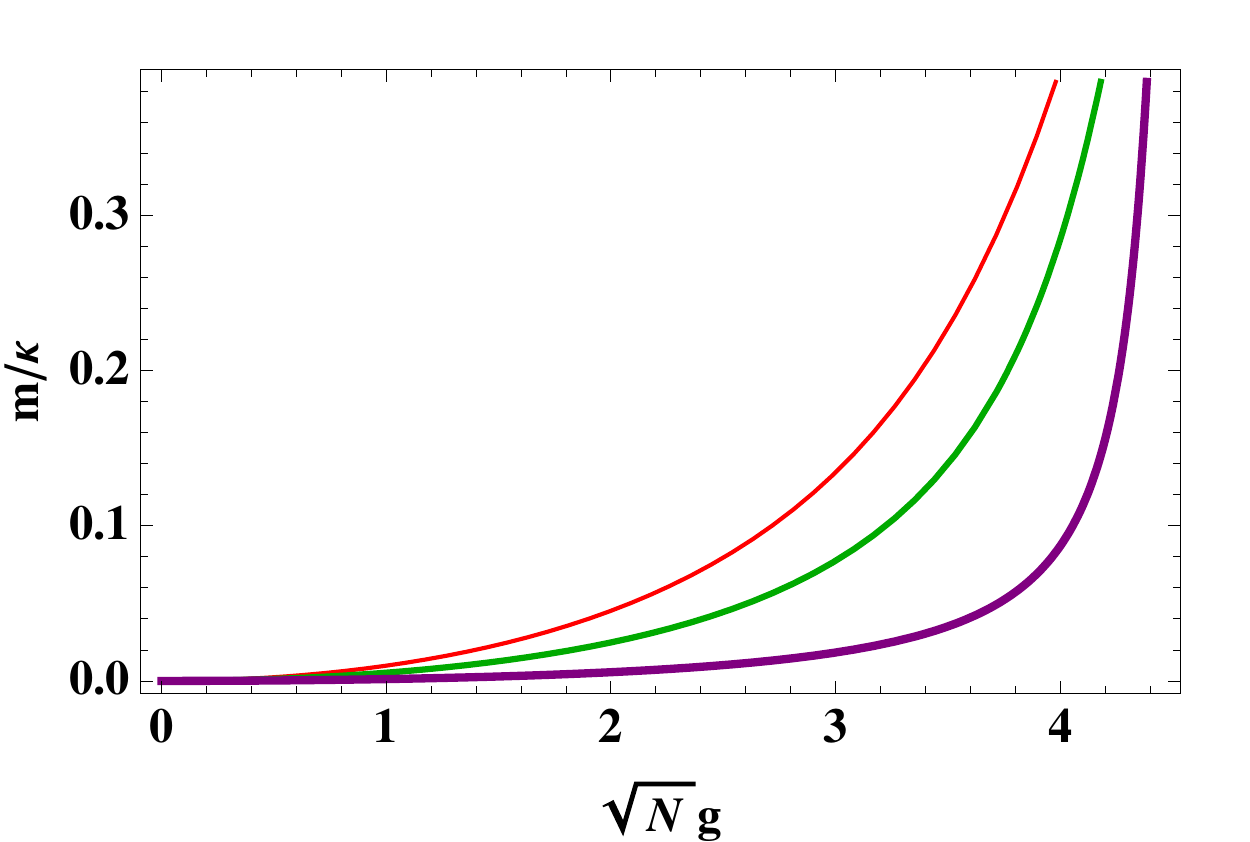}
\includegraphics[width=2.9in]{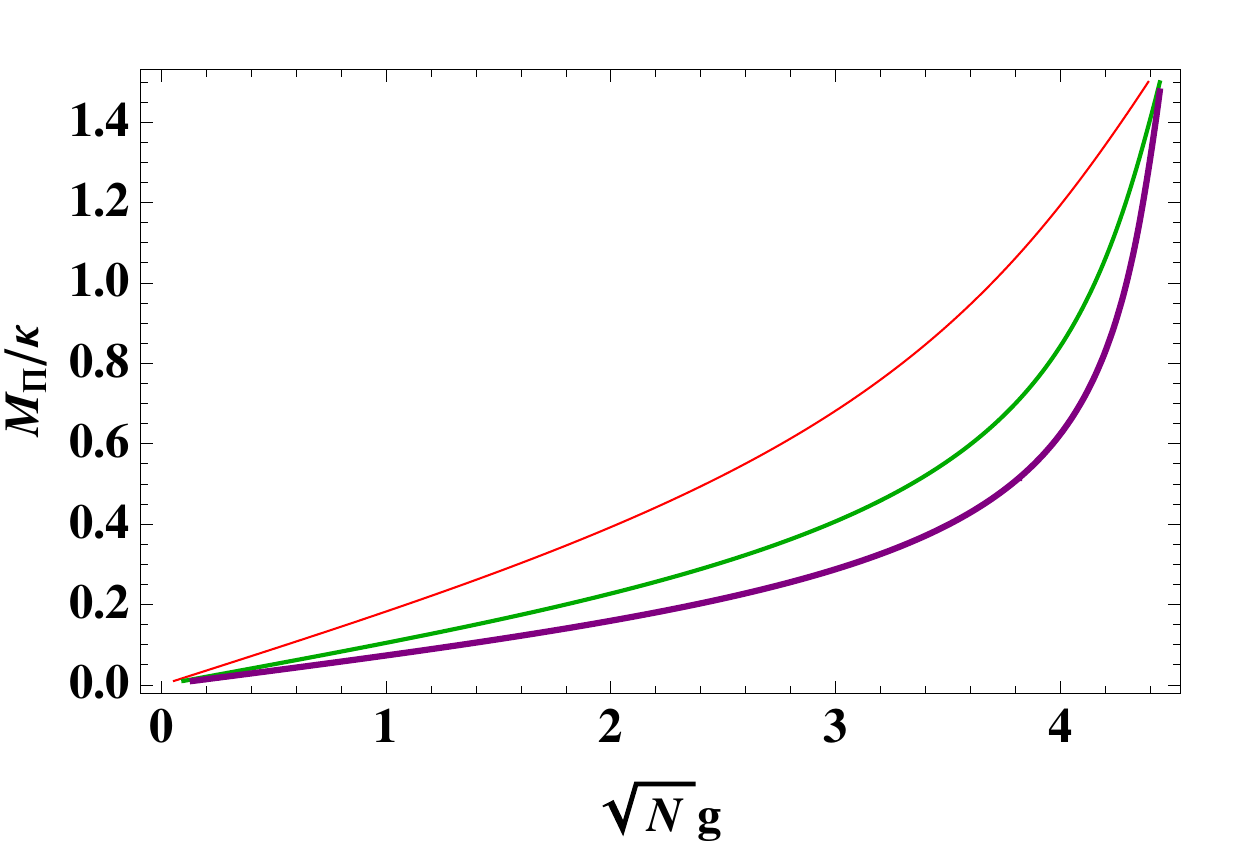}
\includegraphics[width=2.9in]{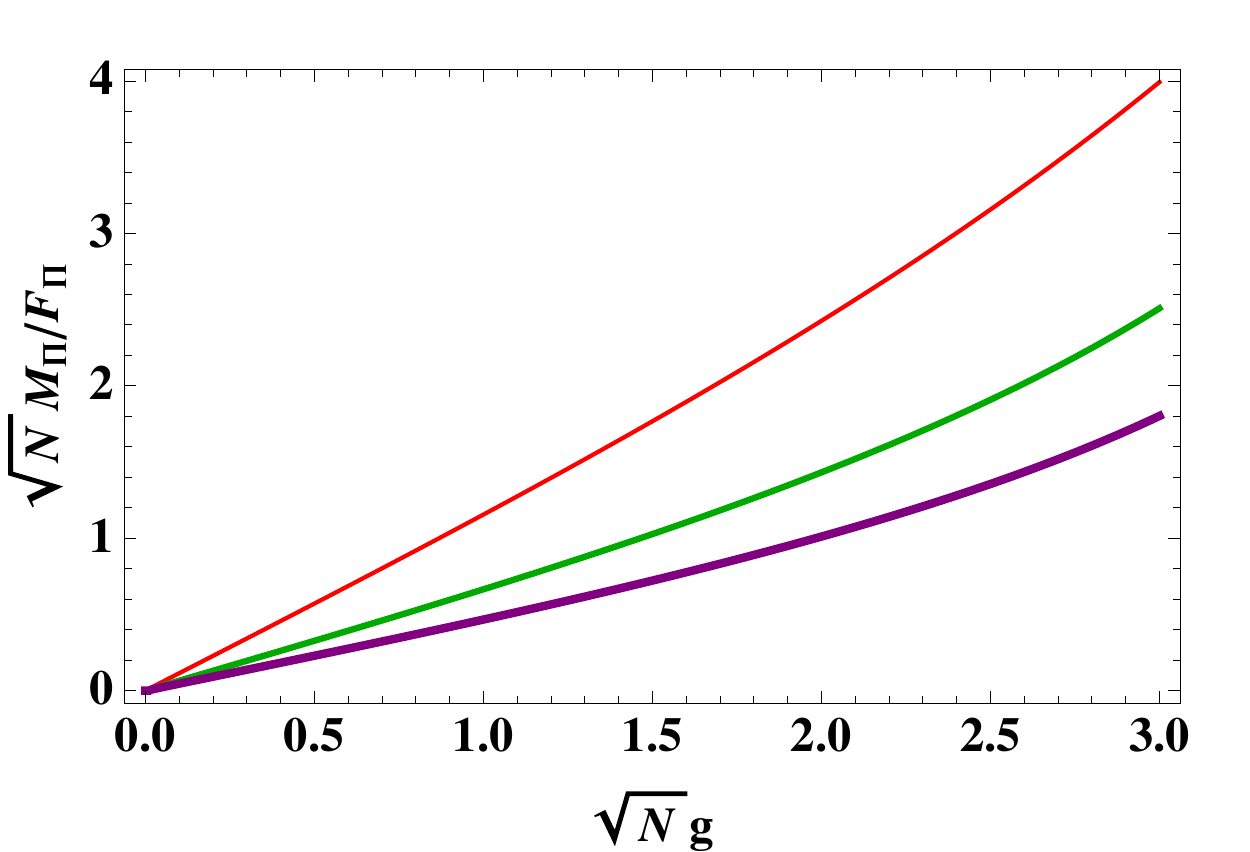}
\includegraphics[width=2.9in]{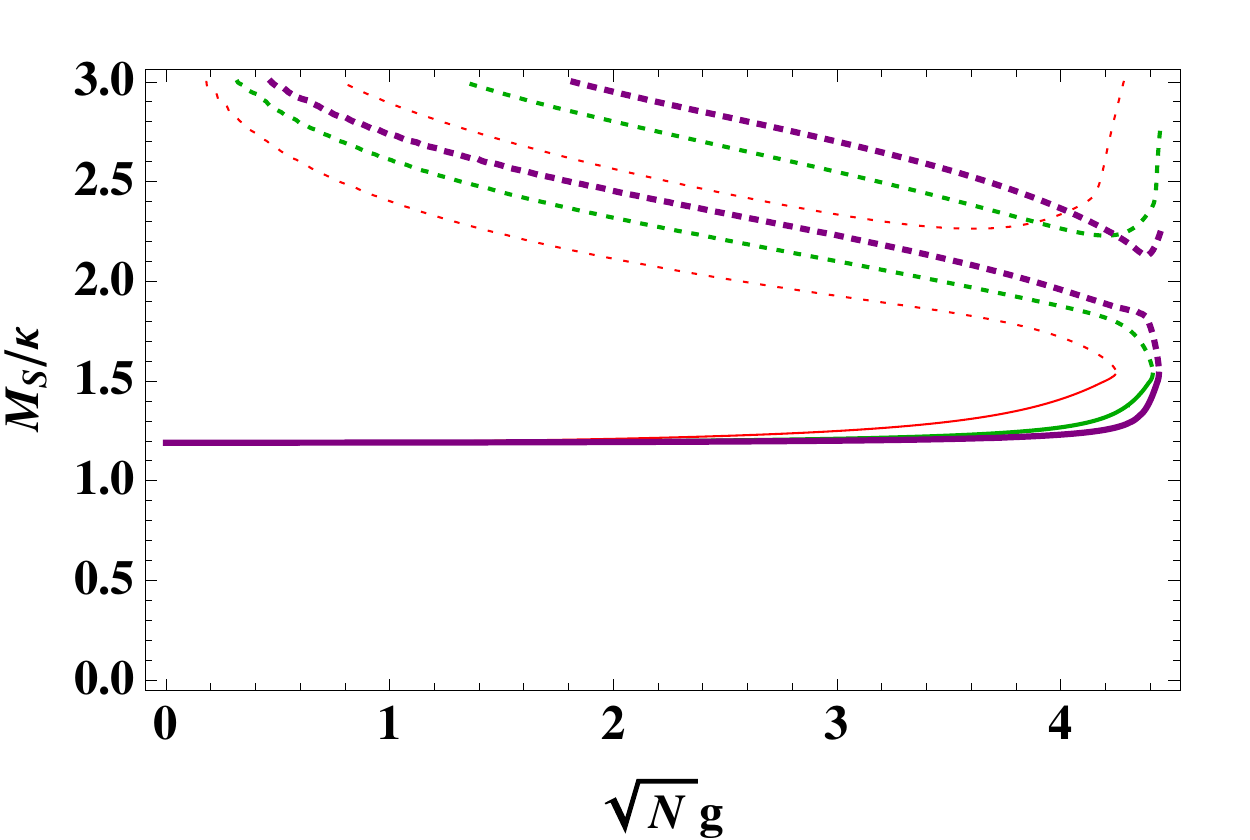}
\caption{Numerical results obtained by solving (\ref{Eq:m}), (\ref{Eq:MassMP}), (\ref{Eq:FP}) and (\ref{Eq:PS}) for $m$, $M_\Pi$, $F_\Pi$, and $M_S$. Each curve corresponds to ${\cal M}/\kappa=4$ (thin red), 7 (medium-thick green), and 10 (thick purple). We note that the chiral-symmetry breaking mass is smaller than the confinement scale for couplings below the critical value (\ref{Eq:crit}). As expected, the pseudoscalar mass grows with $g$, whereas the scalar mass is nearly unaffected by the 4FO. The scalar self-energy features unphysical poles (denoted by dotted lines) which merge with the physical poles in the proximity of the critical coupling. The latter is therefore an upper value above which the model breaks down.}
\label{Fig:Results}
\end{figure}
We now solve numerically the equations (\ref{Eq:m}), (\ref{Eq:MassMP}), (\ref{Eq:FP}) and (\ref{Eq:PS}), to obtain $m$, $M_\Pi$, $F_\Pi$, and $M_S$. We do so by only exploring subcritical 4F couplings, {\em i.e.} couplings below the critical value for fermion condensation via the 4FO \cite{Bardeen:1989ds}:
\begin{equation}
g_c = \sqrt{\frac{2\pi^2}{N}}\ .  
\label{Eq:crit}
\end{equation}
Above this value the series which leads to (\ref{Eq:PSchain}) does not converge, and we need to introduce an additional scalar field to account for condensation. Furthermore we should only consider values of ${\cal M}/\kappa$ which are well above unity, or else 4FOs are no longer useful approximations for the underlying gauge theory. Yet, ${\cal M}/\kappa$ should not be too large, in order for the 4FO to have sizeable effects on the resonances.

The numerical results are shown in Fig. \ref{Fig:Results} as a function of $\sqrt{N}\, g$, where each curve corresponds to ${\cal M}/\kappa=4$ (thin red), 7 (medium-thick green), and 10 (thick purple). In the top-left figure we plot $m/\kappa$, which turns out to be below unity for subcritical 4F couplings. In the top-right figure we plot $\sqrt{N}\, M_\Pi/\kappa$, an expression of the pseudoscalar mass which does not scale with $N$. Note that for $\sqrt{N}\, g\lesssim 3$ the relationship is nearly linear. Below this value chiral-symmetry breaking is small enough that there is yet an approximately conserved chiral current, and the pseudoscalar decay constant can be meaningfully computed using (\ref{Eq:DefFP}). This gives a result which is approximately independent of 
$g$:
\begin{equation}
F_\Pi \simeq 1.6\, \sqrt{N}\, \kappa\ .
\end{equation}
In this region $\sqrt{N}\, M_\Pi/F_\Pi$ is shown in the bottom-left figure. Note that the latter can be directly tested by lattice computations, as the quantity on the $y$-axis does not involve the confinement scale $\kappa$.

Finally, in the bottom-right figure $M_S/\kappa$ is shown by the solid curves. In the scalar channel there are unphysical solutions to (\ref{Eq:PS}), which are shown by the dotted curves. These are present, at larger time-like momenta, also for the theory in isolation, {\em i.e.} $g\to 0$. A nonzero $g$ brings these unphysical poles to lower momenta. In the proximity of the critical coupling the unphysical and physical poles merge, and the model breaks down. We thus expect this computation only to be reliable for couplings below the critical value. We note that the mass of the lightest scalar is approximately unchanged, as $g$ grows. This would not be unexpected in a local NJL model: in fact the latter can be expressed in an equivalent form as a local 4FO, and the additional 4FO in (\ref{Eq:L}) has the only effect of renormalizing the former, with no implications for the scalar mass. We see that the same result holds approximately for nlNJL models as well, and is a consequence of $S$ and $\Pi^a$ forming a linear multiplet of the chiral symmetry. If, on the other hand, the latter is non-linearly realized, and the scalar $S$ is a singlet of the stability group $SU(2)_V$, large and negative corrections to $M_S$ are possible. However the accuracy of prediction (\ref{Eq:msigma}) suggests that the nlNJL model is a reliable approximation of QCD at energies near $\kappa_{\rm QCD}$.
\section{Discussion}
In this note we have analysed the effect of a 4FO on the mass spectrum of a confining SCGT with two fermions in an $N$-dimensional representation of the gauge group. This SCGT features an $SU(2)_L\times SU(2)_R$ chiral symmetry which is broken by the vacuum state to $SU(2)_V$. The confining sector is modelled by a nlNJL action with confinement scale $\kappa$, where the latter is the mass of the confined fermions at zero momentum. The 4FO is chosen to violate chiral symmetry, while preserving $SU(2)_V$. We only analysed the behaviour of the lightest pseudoscalar resonance, {\em i.e.} the pNGB isotriplet $\Pi^a$, and the lightest scalar resonance, the isosinglet $S$. We did the computation in the large-$N$ limit, which allowed us to consider both perturbative and nonperturbative values of the 4F coupling $g$, although we required the latter to always be below the critical value $g_c$ for condensation.

After checking that this model properly accounts for the experimental QCD data on the pions and the $\sigma$ meson, we computed $M_\Pi$ and $M_S$. As expected, we find that $M_\Pi$ grows with $g$, as the latter violates chiral symmetry and gives a positive symmetry-violating fermion mass $m$. On the other hand, the lightest scalar resonance receives no large corrections to its mass, which turns out to sightly grow from $M_S/\kappa\simeq 1.2$ at $g=0$ to $M_S/\kappa\simeq 1.5$ at $g\simeq g_c$ . As the 4F coupling approaches the critical value, unphysical solutions merge with the physical poles, and the model breaks down. 

This type of computation can be applied to theories of DEWSB in which the Higgs boson is itself a pNGB. An example is provided by the recent theories of fundamental composite dynamics \cite{Cacciapaglia:2014uja,Arbey:2015exa}. In this case, since $\kappa$ is expected to be several hundreds of GeV or even heavier, we would likely need symmetry-violating 4FOs with couplings well below the critical value. On the other hand, theories of DEWSB in which the Higgs is the lightest scalar resonance appear to be disfavoured, as the latter is shown to always have a mass of the order of $\kappa$, regardless of whether the coupling $g$ is perturbative or nearly critical. However we expect these results to be only viable for running SCGTs, whereas nlNJL models which are only characterised by a confinement scale are unlikely to properly account for SCGTs with walking dynamics.

Aside from the application to DEWSB, it is the author hope that this type of analysis will also stimulate lattice research, as testable predictions are now made available. For instance, the bottom-left plot of Fig. \ref{Fig:Results} can be directly tested on the lattice. An example of lattice computation which is related to our analysis is provided by the study of gauged NJL models \cite{Catterall:2011ab,Catterall:2013koa}, with 4FOs which are themselves local NJL Lagrangians. It should be possible to generalise these models to account for explicit chiral symmetry breaking, and evaluate the mass spectrum of the lightest resonances as a function of the four-fermion coupling.
\section*{Acknowledgements}
The author thanks K. Tuominen for useful suggestions.
\bibliographystyle{aipnum4-1}

\bibliography{PseudoGB.bib}

\end{document}